# Service-Dominant Business Model Financial Validation: Cost-Benefit Analysis with Business Processes and Service-Dominant Business Models

**Egon Lüftenegger**

CAMPUS 02 University of Applied Sciences

IT and Business Informatics

Körblergasse 126, 8010 Graz, Austria

egon.lueftenegger@campus02.at

**Selver Softic**

CAMPUS 02 University of Applied Sciences

IT and Business Informatics

Körblergasse 126, 8010 Graz, Austria

selver.softic@campus02.at

**Abstract**. In this paper, we present our software-supported method for analyzing the economic feasibility of business models. The method integrates the business models and business processes perspectives for analyzing how a company appropriates the financial cost and benefits. In this method, we use the Service-Dominant Business Model Radar to specify business models, then translate the specified business model into a business process for analyzing the financial feasibility of a business model. At the final step in our method, we use the generated business process in the previous step with a software-based tool, The Cost-Benefit Tracker, for analyzing the economic potential of the business model. We designed and developed the Cost-Benefit Tracker as a simple software-based BPMN 2.0 tool by integrating the concepts of the Service-Dominant Business Model Radar tightly. As a result, the software is simple and straightforward to use than enterprise BPMN 2.0 software. Hence, entrepreneurs can use the presented software-supported method to financially evaluate business model concepts specified with the Service-Dominant Business Model Radar.

## 1 Introduction

Customers are moving from buying products towards integrated solutions (Vargo & Lusch, 2004). Furthermore, Customers are moving from buying physical goods to digital services as solutions. Therefore, the business model design is shifting from a Goods-Dominant (G-D) perspective towards a Service-Dominant one by adopting a Service-Dominant (S-D) Logic (Lüftenegger, 2014). Under this new logic, the business model concept has been reframed as the Service-Dominant Business Model (Lüftenegger, 2014). The Service-Dominant Business Model takes the value network organizational structure approach of the S-D Logic instead of the traditional value chain approach of the G-D Logic. This organizational structural change is required for designing solutions as value co-creation between business actors such as users and companies. The value co-creation takes places within a business ecosystem: the value network. Furthermore, the rise of digital services requires tools for modeling digital ecosystems as business models (Luftenegger, Comuzzi & Grefen, 2013).

A business engineering framework that combines business strategy, business models, business processes as service compositions, and business services was developed by adopting the Service-Dominant Logic (Lüftenegger, 2014). In prior works, strategy and business models' aspects of the framework were developed as management tools: The Service-Dominant Strategy Canvas (Lüftenegger, 2014; Lüftenegger, Comuzzi & Grefen, 2017) and the Service-Dominant Business Model Radar (Lüftenegger, 2014). In this research work, we present our software-supported method. Our contribution is twofold: First, our method facilitates the financial evaluation of business models by transforming business models into business processes. Second, We developed a software-based business process analysis tool that is highly integrated with the Service-Dominant Business Model Radar. This integration is needed for achieving a mechanism to evaluate business models designed or represented with the Service-Dominant Business Model Radar in terms of financial costs and benefits.

In Service-Dominant Business Models, value is co-created and shared between actors of a value network. By tracking costs and benefits in a business process, we can help entrepreneurs with our software-supported method to understand how the value is shared among the actors of the Service-Dominant Business Model. The value shared among the parties has been explored by using business model tools such as e3-value (Gordijn & Akkermans, 2001). However, a method that shows how the financial costs and benefits by integrating the business model level with the business process level has not been previously developed. Hence, the novelty of our approach. Our method is also relevant because how the value is captured and shared is acknowledged as an essential topic in business model research (Zott & Amit, 2007; Chesbrough, 2010).

In this paper, we present our software-supported method on how value is captured and shared by analyzing business processes derived from business models. Specifically, in our software-supported method, we describe how to transition from business models designed with the Service-Dominant Business Model Radar (BMR) to business processes represented in BPMN 2.0. Then, we use the business process to perform a cost-benefit analysis with our software: The Cost-Benefit Tracker. We achieve quantitative financial evaluation by integrating the specific aspects of the Service-Dominant Business Model Radar into our software.

## 2 Design Research

New entrants into the market are developing new business models to disrupt traditional companies (Lüftenegger, Angelov, Van der Linden & Grefen, 2010; Lüftenegger, Angelov & Grefen, 2011). Motivated by the use of the Service-Dominant Business Model Radar in entrepreneurship, we came with the following research question: How can we financially evaluate business models designed with the Service-Dominant Business Model Radar?

For answering this research question, we use the design science research method. Design *s*cience is suited for the evaluation of conceptual artifacts such as methods and software (Weringa, 2014). We use the design science approach to validate our software-supported method with a case study (Hevner, March & Ram, 2004).

As a starting point, we performed a literature review on the Service-Dominant Business Model Radar (See Section 2). As a result of the literature review, we conclude that an artifact in the form of a software-supported method is suitable for achieving this goal. As first step, we developed the first part of the method to transform the elements of a business model specified with Service-Dominant Business Model Radar into a business process. Then, we developed the second part of our method by designing and implementing a simple software tool to integrate tightly the characteristics of the Service-Dominant Business Model with process models specified in Business Process Modeling Notation (BPMN).

We choose particularly BPMN 2.0 (Object Management Group, 2011) because it is an emerging standard for specifying business processes developed by a wide range of Business Process Modeling (BPM) vendors. This standard is one of the most important forms of business process modeling representation by offering clear semantics to describe the business process of a company (Zor, Leymann & Schumm, 2011; Allweyer, 2016). This language was developed with the intention to model typical business modeling activities (Zur Muehlen & Recker, 2008). This is another important reason for choosing this notation due to the business-orientation of our software-supported method. Our software integrates the concepts of the Service-Dominant Business Model radar to complement the translated specification in BPMN 2.0 for performing financial evaluations of business models. The goal of the software is to provide entrepreneurs a simple BPMN 2.0 tool without the complexities and costs of enterprise software such as IBM WebSphere Business Modeler.

The resulting artifact is a software-supported method with three steps. We evaluate our software-supported method with a use case scenario: The ad-supported business model of a music streaming company. Nowadays, Spotify is a music streaming company that offers a music streaming service (Kreitz & Niemela, 2010). In an ad-supported business model, the focal organization behind a business model offers a product or service for free and gains revenue from advertising (Hanson, 2000; Osterwalder & Pigneur, 2010; Gassmann, Frankenberger & Csik, 2014). We evaluate the three steps of our methods as follows: First, in Section 4.1, we represent the ad-supported business model with the Service-Dominant Business Model Radar for a streaming company like Spotify. Next, in Section 4.2, we use the method to generate a BPMN 2.0 process: A collaboration diagram is a BPMN diagram that features two or more pools (Dumas, La Rosa, Mendling & Reijers, 2018). A collaboration diagram is a BPMN diagram that Finally, in Section 4.3, we use our software to financially analyze the s ad-supported business model with the generated BPMN 2.0 process.

## 3 Background and Related Work

### 3.1 A Framework for Service-Dominant Business Design and Engineering

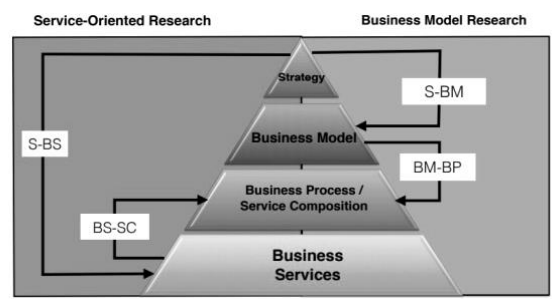

**Figure 1.** Service-Dominant Business Framework and research areas (Lüftenegger, 2014)

The Service-Dominant Business Framework, developed in (Lüftenegger, 2014), is a four-layer approach that integrates: Strategy, Business Models, Business Processes (Service Compositions) and

Business Services. The framework was developed on the foundations of previous works of business design and engineering (Al-Debi, El-Haddadeh, and Avison, 2008; Al-Debei and Avison, 2010; De Castro, Marcos and Wieringa, 2009; Osterwalder and Pigneur, 2002; Sanz et al., 2007). The Service-Dominant Business Framework integrates business model research with service-oriented research. In the service-oriented research aspect of the framework, a modular strategy (S) into business services (BS) enable an organization to achieve agility for reacting to changes in the market such new trends or new customer needs (S-BS in Figure 1). The business service composition (BSC) uses business services (BS) as tasks within a business process (BS-SC in Figure 1).

In the business model research aspect of the framework, the strategy leads to business models (BM) (S-BM in Figure 1). Finally, business models are represented and operationalized as business processes (BM-PB in Figure 1). We explain each of the framework's layers as follows (Lüftenegger, 2014):

**Strategy.** The strategy concept reframed into the Service-Dominant Logic result in a Service-Dominant Strategy. The resulting management tool, Service-Dominant Strategy Canvas, aims to shift the mindset from a Goods-Dominant Logic towards a Service-Dominant Logic.

**Business Models.** The business model concept reframed into the Service-Dominant Logic correspond to a Service-Dominant Business Model. The resulting business model tool is the Service-Dominant Business Model Radar: In short, Business Model Radar (BMR). The Business Model Radar is a tool for designing and analyzing business models as ecosystems. An overview of the BMR is presented in Section 3.2 and an illustrative example is presented in Section 4.1.

**Business Processes (Service Composition).** Business process operationalizes the business models. These business processes are the Business Service Compositions (BSC) of a task or a sub-process that are represented as Business Services.

**Business Services.** Business Services are service modules that can be inside or outside the organization. Business Services outside the organization in combination with business services outside the organization enable the cooperation with external business actors. This collaboration can be represented the Business Process Layer as a Cross-organizational Business Processes (business process across organizations).

In this research document, we concentrate our focus in the relationship between Business Models and Business Processes layers of the Service-Dominant Business Framework in the context of business model research: how the financial value is captured and shared among the business collaborators. In our case, the Co-creators of a Service-Dominant Business Model represented as a Business Model Radar.

## 3.2 The Service-Dominant Business Model Radar: A management tool for designing and analyzing Service-Dominant Business Models

The Service-Dominant Business Model Radar is a management tool for designing and analyzing Service-Dominant Business Models. A Service-Dominant Business model differentiates from the traditional business model by taking a network structure based on the Service-Dominant Logic rather than the value chain approach of the Goods-Dominant Logic (Lüftenegger, 2014).

In (Lüftenegger, 2014), the author develops the Service-Dominant Business Model Radar by combining Design Science with Action Research: Action Design Research. In this work, the author makes an explicit distinction between cost and benefits in the Service-Dominant Business Model for enhancing the visibility and the semantics of each component. Researchers and practitioners applied a simplified version of the management tool presented in (Luftenegger et al., 2013) for conducting workshops sessions. In Section 4, we present our software-supported method by depicting how to use the complete version of the Service-Dominant Business Model Radar (Lüftenegger, 2014) for the translation of elements of a Service-Dominant Business Model into a Business Process. In a Service-Dominant Business Model Radar we can distinguish the following main elements (Lüftenegger, 2014):

**Solution (S).** The solution is the goal of the Business Model. We can think about the solution as the Co-created Value between the Co-Creation Actors of the Business Model.

**Co-Creation Actors (A).** This element represents the participants in a business model collaboration. We can distinguish between User, Focal Organization, and Partners. The Focal Organization is the Co-Creation Actor that is behind the Business Model that uses Partners as co-creators for co-creating a solution with the User Co-Creation Actor.

For each Co-creation Actor defined above, we can distinguish the following four elements:

**Value Proposition (VP).** This element represents the Value Proposition of each Co-Creation Actor into the Solution. Each Actor's Value Proposition can be seen as a part of the overall Solution.

**Co-Creation Activities (CA).** This element represents the activities that a Co-Creation Actor performs for delivering a value proposition into the solution.

**Costs (C).** This element represents what are the costs that a Co-Creation Actor incurs by performing Co-creation activities and/or participating in the business model.

**Benefits (B).** This element represents what are the benefits that a Co-Creation Actor gain by performing Co-creation activities and/or participating in the business model.

As shown in Figure 2, each Co-creation Actor $A_i$ from $A_1$ to $A_N$ is represented as a slice from the BMR. There is only one Solution (S), represented at the center of the BMR. The central position embodies that the solution is co-created by all the Co-creation Actors in the business model.

The design of a business model with the BMR is achieved by answering a set of questions. For the solution at the center of the BMR, we ask the following: What is the value that are we co-creating?

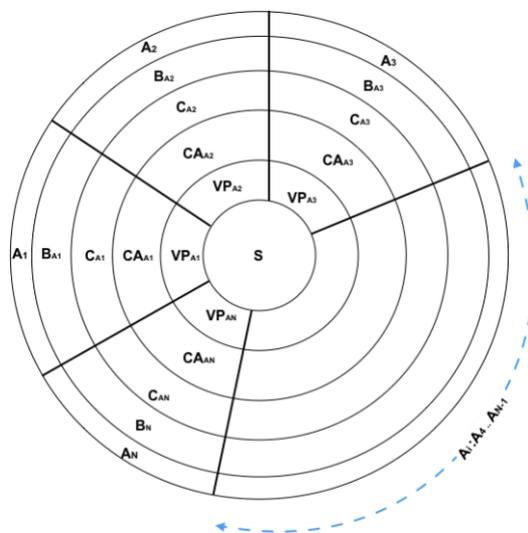

**Figure 2.** Representing Actors in the Service-Dominant Business Model Radar (BMR)

For identifying the co-creation Actors, we answer the following questions: Who is the focal organization, behind the business model? Who is the user? Who are the Partners? We represent each actor as a slice in the BMR.

After the co-creating actors are identified and labeled in the BMR by filling the names of each Co-creating Actor $A_i$ from $A_1$ to $A_N$, we answer the following questions for each Co-creating actor $A_i$: What is the Value Proposition that the co-creation actor brings in the solution? We place the answer in $VP_{Ai}$. What are the co-creating activities that a Co-creator actor performs to provide the value proposition? We place the answer in $CA_{Ai}$. What are the costs that an actor incur? We place the answer in $C_{Ai}$. What are the benefits that an actor gain? We place the answer in $B_{Ai}$. In Section 4.1, we use this question to present an illustrative example.

# 4 Service-Dominant Business Model Financial Validation: integrating business concepts with business processes at a financial level

The Service-Dominant Business Model is useful for analyzing and designing business concepts from an ecosystem perspective (Lüftenegger et al., 2013; Lüftenegger, 2014). These business concepts require financial validation to assess its viability. Developing the complete business model for testing it in the market requires a great number of financial resources. A method to simulate the validity of the business model concept could potentially be a safe way of analyzing the financial feasibility of the business idea before committing financial resources for a real test in the market. Many investors and entrepreneurs got burned during the dot com crash due a large amount of capital invested in flawed business models (Magretta, 2002). Several of these performance evaluations are performed form the qualitative rather than a quantitative aspect of business models (Malone, Weill, Lai, D'Urso, Herman, Apel & Woerner, 2006).

We propose to perform financial validation of business models concepts as a three steps software-supported method: The Service-Dominant Business Model Financial Validation method. We define our software-supported method as the following three steps:

**Step 1.** Represent a business model with the Service-Dominant Business Model Radar (See Section 4.1)

**Step 2.** Transform the specified business model in Step 1 to a business process represented in BPMN 2.0 collaboration diagram (See Section 4.2) by following steps 2.1 to 2.5.

**Step 3.** Use our Cost-Benefit Tracker software with the business model represented as a process in BPMN 2.0 in Step 2 to determine the costs and benefits of the business model (See section 4.3)

## 4.1 Representing the Business Model with the Business Model Radar

We explain how the Service-Dominant Business Model Radar works with an illustrative example within the context of Spotify's freemium Business Model (Wagner, Benlian & Hess, 2014). In this business model, we have a combination of two business models: the ad-supported business model and premium business model. In this research paper, we use the ad-supported business model. We choose this business model because it is how the freemium business model reaches the market for the first time. Hence, the viability of the free aspect is critical for the success of the business model concept. Once, the ad-supported business model works, it can evolve into a complete freemium business model by integrating the premium business model (Pauwels & Weiss, 2008). For generalizing the ad-supported business model with other companies, we use the name Streamer to refer to Spotify or other streaming companies using this business model.

In the Streamer´s ad-Supported business model, a user obtains free streaming music financed by advertising interruptions. In Figure 3, we represent the ad-supported music streaming business model with the BMR by using the template of Figure 2 and answering the questions of Section 3.2, we identify four Co-creation actors: The Free User as User (U1), Streamer as the Focal Organization (FO), and, the Record Label as Partner 1 (P1) and the Advertiser as Partner 2 (P2). These actors are the co-creators of the ad-supported music streaming solution as the business model.

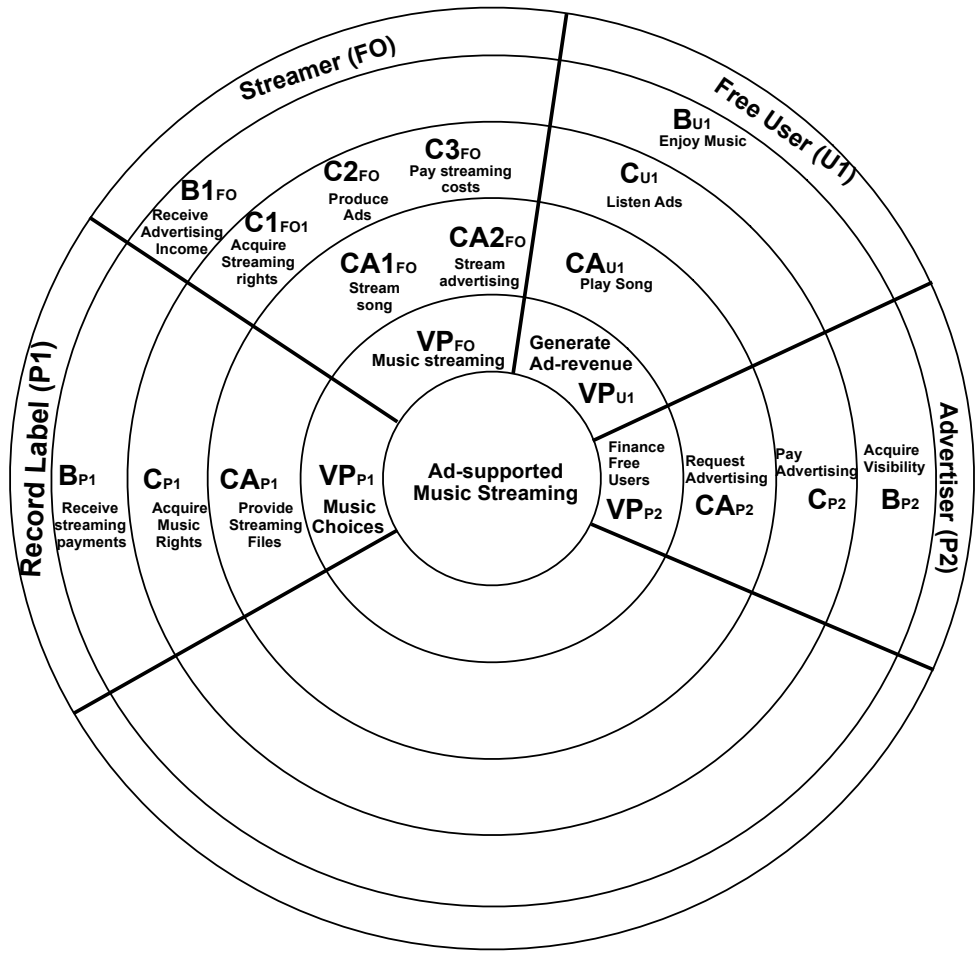

**Figure 3.** Ad-supported music streaming business model represented with the Service-Dominant Business Model Radar (BMR)

The Streamer plays the role of the Focal Organization that is behind this business model. For delivering the service to the user, they require the combination of different value propositions by Co-creating Actors: "music choices" $VP_{P1}$ by Record Label, "finance free users" $VP_{P2}$ by Advertiser, "generate advertising-revenue "$VP_{U1}$ by Free User, and "music streaming" $VP_{Fo}$ by Streamer. These Value Propositions are represented in the BMR shown in Figure 3.

Each Value Proposition in the BMR is achieved by a Co-creation Activity that has Costs and Benefits. We explain each of them in the ad-supported music streaming business model as follows: The "music choices" $VP_{P1}$ is enabled by the "provide streaming files" $CA_{P1}$. This $VP_{P1}$ has "receiving streaming payments" as Benefit $B_{P1}$ and "acquire streaming rights" as cost $C_{P1}$. The "music streaming" $VP_{Fo}$ is enabled by two Co-Creation Activities: "stream song" $CA1_{FO}$ and "stream advertising" $CA2_{FO}$. The "stream song" Co-Creation Activity $CA1_{FO}$ has "acquire streaming rights" $C1_{FO}$, "produce ads" $C3_{FO}$ and "pay streaming costs" $C2_{FO}$ as Costs. The "stream advertising" $CA2_{FO}$ Co-creation Activity has "produce ads" $C2_{FO}$ as Cost and "receive advertising income" $B1_{FO}$ as Benefit.

The Free User "U1" and the Advertiser "P2" are key for having a working business model. The Advertiser finances the Free User by requesting advertising to the Streamer. This action generates a cost "pay advertising" $C_{P2}$ and a benefit "acquire visibility" $B_{P2}$ for the Advertiser. The Free User "U1" is essential for generating revenue to Streamer FO: The "generate Ad-revenue" $VP_{U1}$ is achieved by the "play song" $CA_{U1}$. This revenue generation happens because the Free User listen ads as Costs represented by "listen ads" $C_{U1}$.

All the Cost and Benefits, represented in the BMR, for each Co-Creator Actor explain what each of them gains and losses by being part of the co-created solution: The ad-supported music streaming business model.

## 4.2 Transforming Service-Dominant Business Model elements into a BPMN Collaboration diagram

As presented in (Lüftenegger, 2014), the Business Process layer bridges the Service-Dominant Logic and Service-Oriented perspectives on Service by putting a focus on the customer during the composition of business services. Methods that follow this approach for defining business processes from Service-Dominant Business Models have been proposed (Lüftenegger, 2014; Suratno, Ozkan, Turetken, & Grefen, 2018). In particular, in (Suratno et al., 2018), the approach on service composition is achieved only for the execution of business processes into a Business Process Management System (BPMS). However, the business model aspects of a Service-Dominant Business Model are forgotten due to the omission of costs and benefits that are the motivation drivers of a focal organization in developing and executing a business model.

In (Di Valentin, Burkhart, Vanderhaeghen, Werth, & Loos, 2012), the authors propose the identification of the organization (process owner and process participants), implementation (process and information systems), performance flow (input and output) and information flow (input and output). By translating our Co-Creation Actors into a business process, we can identify as the Focal Organization as the process owner that collaborates with the User and the Partners: the process participants. The performance flow is in our case is driven by the financial performance of a Business Model with costs and benefits as inputs and outputs. The Information Flow is driven by the Co-Creation Activities that need to be integrated with the previously identified Performance Flow. By refining and adapting the methods presented in (Lüftenegger, 2014; Suratno et al, 2017), we propose the following method:

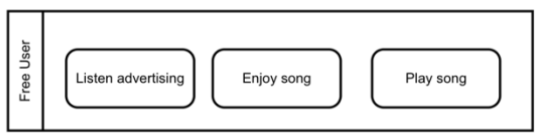

**Figure 4.** Step 2.1: User Actor Pool for Free User

**Step 2.1.** Identify the User Co-creator Actor from the BMR and place it as User Pool at the top of the BPMN 2.0 collaboration diagram. Identify the Costs, and Benefits of each Co-creation Activity for each User Co-creation Actor in the BMR and place the Costs, Co-creation Activities and Benefits as tasks in the corresponding User Actor Pool in the BPMN 2.0 collaboration diagram in the following order: Cost, Co-creation Activity, Benefit. For instance, in Figure 4, the User Actor Pool identified from the Free User "U1" Co-creation Actor of the BMR shown in Figure 3.

**Step 2.2.** Identify the Focal Organization Co-creator Actor from the BMR and place it as Focal Organization Pool in the BPMN collaboration diagram, below the User Pool. Identify the Costs and Benefits of each Co-creation Activity for the Focal Organization Co-creation Actor in the BMR and place the Costs, Co-creation Activities and Benefits as tasks in the corresponding Focal Organization Actor Pool under the User Actor Pool in the BPMN 2.0 collaboration diagram in the following order: Cost, Co-creation Activity, Benefit. For instance, in Figure 5, the Streamer (Focal Organization) is identified from the BMR of Figure 3 and placed below of the Free User (User) Pool.

**Step 2.3.** First, identify the Partners Co-creators Actors from the BMR and place each of them into Pools with their respective names below the Focal Organization Pool. Next, identify the Costs and Benefits of each Co-creation Activity of each Partner Co-creation Actor in the BMR and place the Costs, Co-creation Activities and Benefits as tasks in the corresponding Partner Actor Pool under the Focal Organization Pool in the BPMN 2.0 collaboration diagram in the following order: Cost, Co-creation Activity, Benefit. For instance, the Advertiser and Record Label in Figure 6 identified from the BMR shown in Figure 3.

**Step 2.4.** Re-organize the order of Tasks if needed by logically connecting the identified elements Cost, Co-creation Activities and Benefits within the Pools of the BPMN 2.0 collaboration diagram. In figure 7, we illustrate Step 2.4 as follows: In the ad-supported Music Streaming BMR, the Free User first listens to an ad; then the song is streamed to the user. In this case, the logic works as the current order of tasks within the Free User Pool of the BPMN 2.0 collaboration diagram. Then we establish a sequential flow by connecting tasks in the diagram.

In the Streamer Pool, the first three tasks Produce advertising, Stream advertising and Receive advertising income make sense as sequence from the BMR, hence we connect them. However, sequence of Pay streaming costs, Stream song and Acquire streaming rights does not follow the logic of the BMR because the Streamer needs to first acquire the streaming rights of a song for being able to stream a song. So, after the song is streamed, the streamer needs to pay streaming songs. Hence, we reorder these three tasks and connect them to the previously identified tasks.

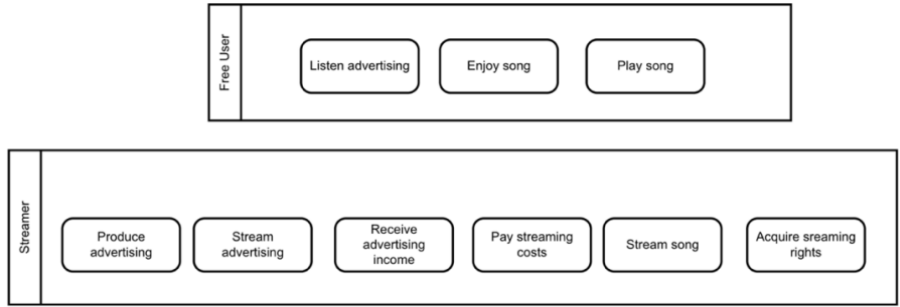

**Figure 5.** Step 2.2: Streamer a Focal Organization Actor Pool placed below the User Pool

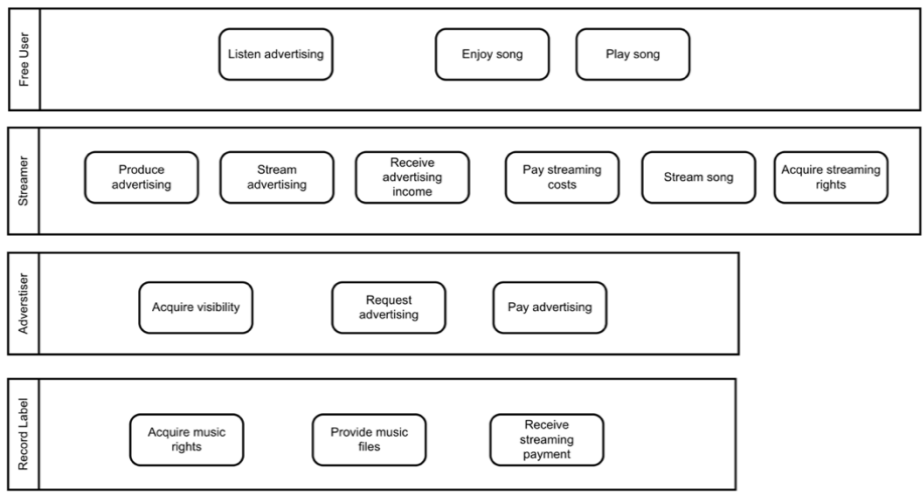

**Figure 6.** Step 2.3: Advertiser and Record Label, the partners of the focal organization, placed below the Streamer (Focal Organization) Actor Pool

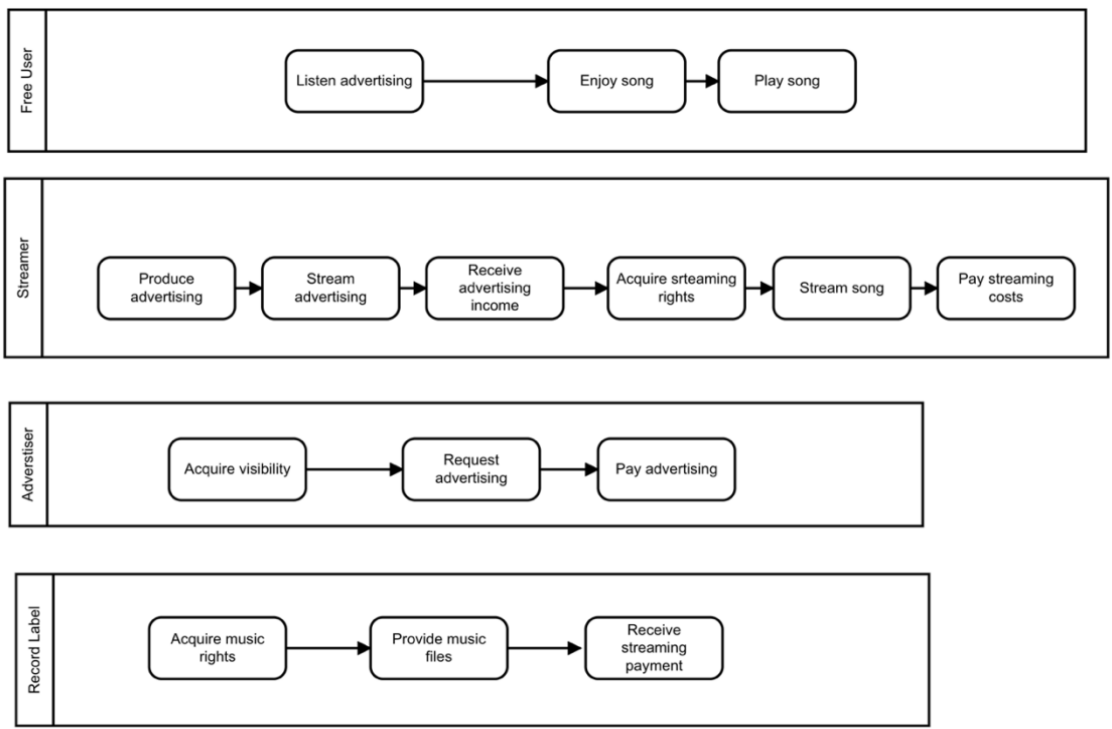

**Figure 7.** Step 2.4: Reorganization and intra-pool connection of tasks

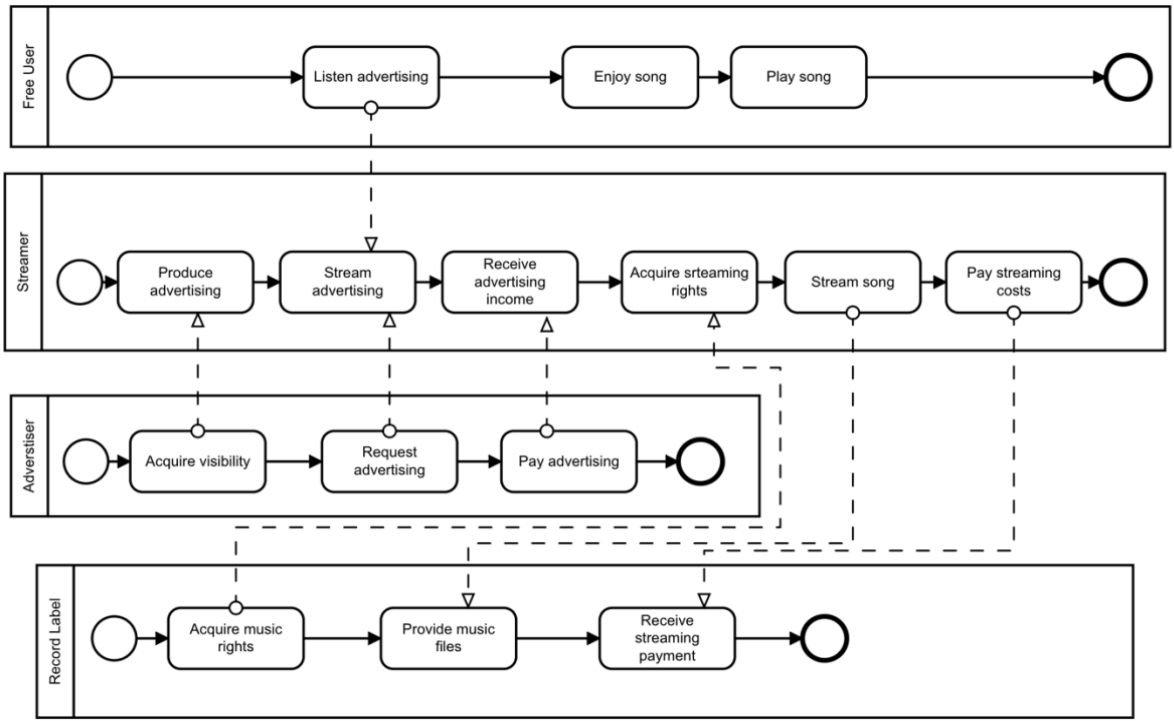

**Figure 8.** Step 2.5: Ad-supported Music Streaming Business Model as Business Process.

**Step 2.5.** Re-organize the flow of Tasks if needed by connecting logically the identified Cost, Co-creation Activities and Benefits tasks across the pools of the BPMN 2.0 collaboration diagram by following the reasoning of the business model represented with the BMR. Finally, add Start and Finish Tasks to the BPMN 2.0 collaboration diagram.

By applying the final of our method (Step 2.5), we have the resulting business process presented in Figure 8. For instance, a logical connection is made between the "listen advertising" User's task and the Streamer's Task "stream advertising". This connection is made because, the "stream advertising" (a Streamer's Co-creation Activity in the BMR) is required by the user for "listen advertising" (a User's cost in the BMR). The user's "play song" Task (a User's Co-creation Activity in the BMR) needs the "stream song" task (a Streamer's Co-creation Activity in the BMR). Hence a connection is made. A connection is made between the "acquire visibility" Task of the Advertiser and "produce advertising" Task of the Streamer, because the acquisition of visibility from the advertiser (a benefit) requires the "produce ad" Task (a co-creation activity). Another connection is the Advertiser's "request advertising" Task (a Co-creation Activity in the BMR) with the stream advertising" Task (another Co-creation Activity in the BMR) because a "request advertising" triggers a "stream advertising" event. The "Receive advertising income" Task (a Cost in the BMR) requires the money flow from the Advertiser with the "pay Advertising" Task (a Benefit in the BMR). Hence a connection is established. Finally, regarding the tasks connections of the Record Label Actor, we have the following: "Acquire streaming rights" Task (a Streamer's cost in the BMR) requires "Acquire music rights" Task (a Record Label's Cost in the BMR). The "pay streaming costs" Task (a Streamer's Cost in the BMR) is connected to the "receive streaming payment" Task (a Record Label's Benefit), because the payment of the streaming costs triggers the receive streaming payment event. Finally, we add the start and finish tasks for each Actor's Pool (in case it applies).

### 4.3 Tracking of Costs and Benefits in BPMN: The Cost-Benefit Tracker

We developed, the Cost-Benefit Tracker (also denoted as "CB Tracker") as a prototype software that uses business processes for analyzing the flow of costs and benefits modeled with the Service-Dominant Business Model Radar. We decided to develop the presented tool instead of using a traditional BPMN 2.0 tool, because we wanted to support the presented method by offering a ready-to-use tool that is highly integrated with the elements of the BMR.

We implemented the Cost-Benefit tracker by using a BPMN 2.0 open source library (see: https://github.com/bzinchenko/bpmnview) because our tool uses a BPMN 2.0 collaboration diagram as input to perform the business model financial analysis. The mentioned BPMN 2.0 library has been tested with a broad set of compatible BPMN 2.0 tools and used in the KPIs (Key Performance Indicators) tracking in Industry 4.0 scenarios (Lüftenegger, Softic, Hatzl &

Pergler, 2018). The full list of compatible modeler tools is available at http://bpmn-miwg.github.io/bpmn-miwg-tools/.

We integrate the Cost-Benefits Tracker with the Service-Dominant Business Model Radar tightly by defining a set of elements associated with each task of a BPMN 2.0 process. We explain each element as follows:

**Actor.** This element associates the owner of the current task within a business process.

**Type.** This element specifies the type of Task is associated with the costs, benefits and, co-creation activities elements of the BMR.

**Goal.** This element specifies the desired achievement of the KPI. This element is useful to group different tasks into a specific goal. For instance, a Goal in a task could be "Profitability."

**KPI.** This element defines the metrics that are we measuring the costs and benefits.

For instance, we can measure the "number of streamed songs" in the Free Music Streaming Business Model.

**Current value.** This element shows the present value of the KPI. The value is represented as a number or a simple formula.

**Target Value.** This element shows the desired value of the KPI. The value is represented as a number or a simple formula.

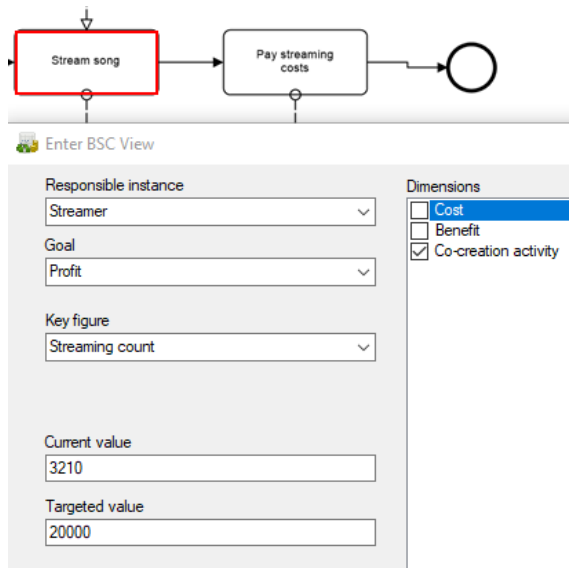

**Figure 9.** Task associated values in CB Tracker

In Figure 9, we can see the achieved integration between the BMR and the business process' tasks. As shown in Figure 9, each task of a Business Process represented with the BPMN 2.0 Notation has an associated Actor, Type, Goal, KPI, Current Value, and Target Value.

In the financial feasibility analysis conducted with the Cost-Benefits Tracker for the ad-supported music streaming BMR (See Figure 3), we have three actors represented in pools within a BPMN 2.0 collaboration diagram: The Users, the Streamer and the Advertiser. The User is the actor in which the user experience is delivered. The Streamer is the actor that takes the role of a Focal Organization (the company behind the business model). The focal organization is the actor that offers the customer experience to the user. Hence, we focus on the financial performance of the Streamer. The Advertiser and Record Label are the Partners that the Streamer needs to make this business model feasible.

The Streamer wants to achieve profitability as its goal. This goal depends on the amount of financial income and financial outcome. In the Free Music BMR, the financial outcome depends on the streamed songs and the advertising income. Hence we can identify in the Streamer Actor's Pool two tasks that are associated with the streamer's financial outcome ("Stream song" and "Pay streaming costs") and two tasks that are associated with the streamer's financial income ("Stream advertising" and "Receive advertising income"). In the Streamer's financial outcome, each streamed song needs to be paid to the Record Label. In the Streamer's financial income, each time that an advertisement (Ad) is streamed by the Streamer, the Advertiser partner needs to pay to the Streamer Actor.

We use the Cost-Benefit Tracker software to measure the financial performance of the Streamer Actor in the Free Music BMR as follows:

**"Stream song" Task (financial outcome).** In the Free Music BMR, a key performance indicator is the number of streamed songs because it is associated with the financial outcome of the Streamer Actor: the Focal Organization that is behind this business model.

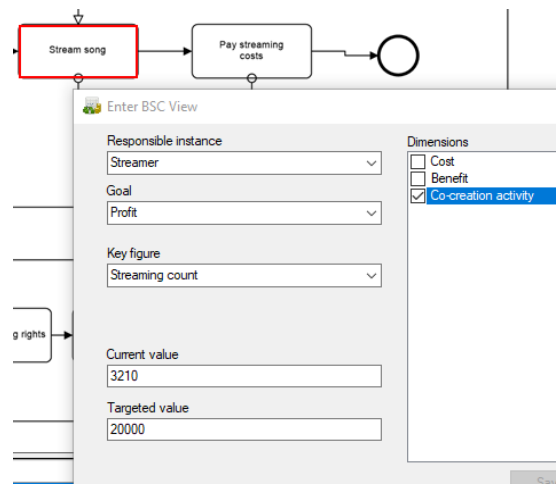

**Figure 10.** Cost-Benefit Tracking for "Stream song" Task in BPMN 2.0 diagram

As shown in Figure 10, we identified the "Stream song" Task and classified as a Co-creation activity type from the BMR. We also associated this task to the Streamer Actor. This task is suitable for a KPI association because we can count the number of streamed songs in the business process. Hence, we establish "Streaming count" as KPI. We can set current the value and the target value of the KPI to 3210 and 20000 (Songs streamed) respectively.

**Pay Streaming Costs Task (financial outcome).** The values identified in the previous task (Stream song) are useful for calculating the KPI associated with this task: Cumulative Streaming (See Figure 11). We can use the KPI associated with the previous task in the business process by calling the task ("Stream song") that includes the value of the KPI. Hence, we can calculate the value of the cumulative streaming KPI by using a formula: "(1.5,Streaming count)*0,45" where 1.5 represents the task id from the previous task which is "Stream song" and "Streaming count" represent the belonging KPI from the same task we want to use. A target value of this KPI is defined: 10000 (euros).

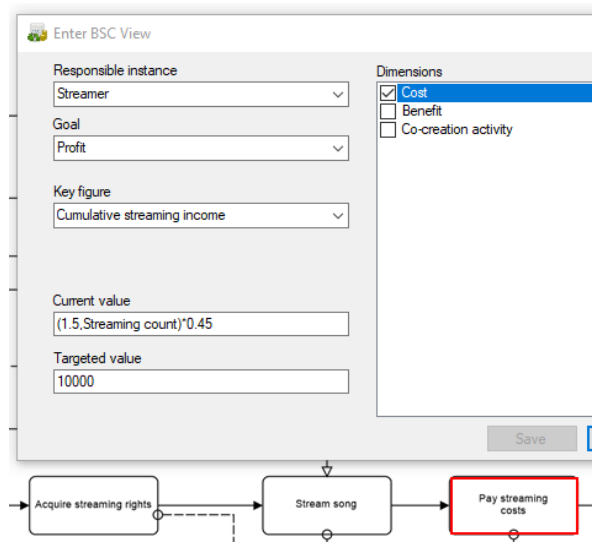

**Figure 11.** Cost-Benefit Tracking for Pay Streaming Costs Task in BPMN 2.0 diagram

**"Stream advertising" (financial income).** This task is important for measuring financial income because we need to know the number of streamed Ads. This number is needed for calculating the income from advertising in the next task of the process ("Receive advertising income" Task). In Figure 12, we show how to use the Cost-Benefit Tracker for the "Stream advertising" Task. This task is associated with the Streamer Actor, classified as a co-creation activity (from the BMR) and "Streaming count" as KPI. The current value of this KPI is defined to 12342 and the target value to 40000.

**"Receive advertising income" Task (financial income).** In Figure 13, we identify the "Receive advertising income" Task as a benefit type (from the BMR). This task can be associated with a KPI for measuring the financial income: "Receive advertising income". In the current value, we define the following formula: "(1.2, Streaming count)*0.5". This formula calculates the number of advertising streaming "Streaming count" from the previous task ("Stream advertising" Task with id 1.2) with a value of 0.5 euros. We can also set the desired value of the formula for achieving profitability.

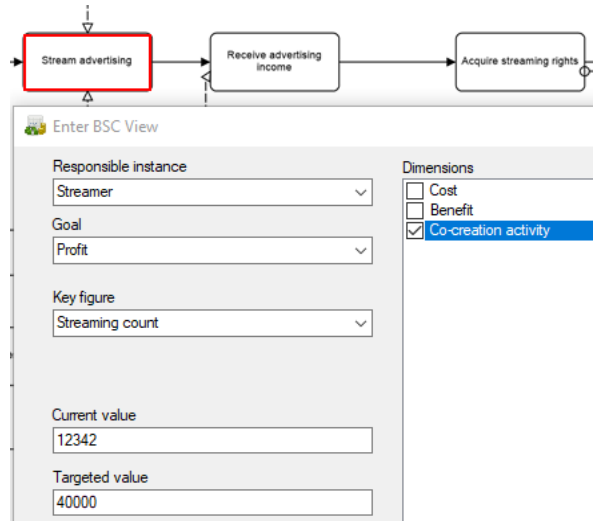

**Figure 12.** Cost-Benefit Tracking for "Stream advertising" Task in BPMN 2.0 diagram

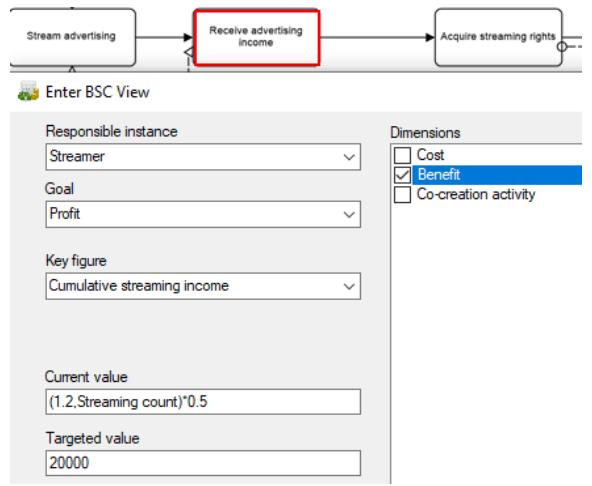

**Figure 13.** Cost-Benefit Tracking for Receive Advertising Income Task in BPMN 2.0 diagram

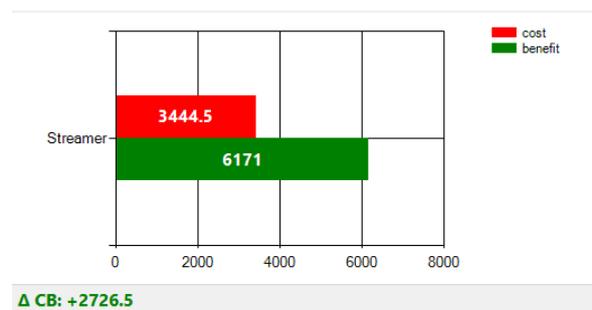

**Figure 14.** Cost-Benefit Overview Diagram for Streamer Actor

The financial feasibility of a business model concept designed with the BMR is proved by the visualization of the total costs and benefits of the identified tasks in the business process: the cost-benefit overview. The cost-benefit overview diagram shown in Figure 14 represents the blueprint that defines how the value is appropriated by the focal organization behind the business model (Johnson, Christensen and Kagermann, 2008). This view is important for the decision-makers within a company for validating the financial viability of a business concept.

# 5 Conclusions

In this paper, we present a software-supported method for analyzing cost-benefits in business models designed with the Service-Dominant Business Model Radar. In a nutshell, we use the presented method to translate business models specified with the Service-Dominant Business Model Radar, and the software tool evaluates the translated business models from the financial perspective of cost and benefits. We developed this method as three steps: At the first step, we represent a business model with the Service-Dominant Business Model Radar. At the second step, we translate the designed business model into a business process by using BPMN 2.0. At the final step, we use our self-developed Cost-Benefit Tracker tool for analyzing the costs and benefits of the business model by using the generated process from the previous step.

The presented software-supported method is novel by two main reasons: First, it is a tightly integrated approach to represent business models as business processes. Second, the method tightly integrates the business models and business process layers of the Service-Dominant Business Model framework at the financial perspective. We developed the software tool named Cost-Benefit Tracker because is a simple and ready-to-use BPMN 2.0 tool that is tightly integrated with the elements of the Service-Dominant Business Model Radar. With the Cost-Benefit Tracker, we can follow how the value is shared between different actors of the business model by providing an operational perspective.

As future work, we will further automate our method for enhancing the usability to non-engineer users. We will integrate BPMN 2.0 modeling capabilities to minimize the complexities introduced in the usage of external BPMN 2.0 modeling tools aimed to process engineers. Regarding the use cases, we will further test our method with a broader set of business models.